\documentclass[twocolumn,showpacs,preprintnumbers,amsmath,amssymb,superscriptaddress]{revtex4}

\usepackage{graphicx}
\usepackage{dcolumn}
\usepackage{bm}

\begin{document}


\title{Rigid-Band Shift of the Fermi Level in a Strongly Correlated
  Metal: Sr$_{2-y}$La$_{y}$RuO$_{4}$}

\author{N. Kikugawa}
\affiliation{School of Physics and Astronomy, University of
  St.~Andrews, St.~Andrews Fife KY16 9SS, United Kingdom} 
\affiliation{Venture Business Laboratory, Kyoto University, Kyoto
  606-8501, Japan} 
\affiliation{Department of Physics, Kyoto University, Kyoto 606-8502,
  Japan} 

\author{A.P. Mackenzie}
\affiliation{School of Physics and Astronomy, University of
  St.~Andrews, St.~Andrews Fife KY16 9SS, United Kingdom} 

\author{C. Bergemann}
\affiliation{Cavendish~Laboratory,~University~of~Cambridge,~Madingley~Road,
  Cambridge~CB3~0HE, United Kingdom}

\author{R.A. Borzi}
\affiliation{School of Physics and Astronomy, University of
  St.~Andrews, St.~Andrews Fife KY16 9SS, United Kingdom} 

\author{S.A. Grigera}
\affiliation{School of Physics and Astronomy, University of
  St.~Andrews, St.~Andrews Fife KY16 9SS, United Kingdom} 

\author{Y. Maeno}
\affiliation{Department of Physics, Kyoto University, Kyoto 606-8502,
  Japan} 
\affiliation{International Innovation Center, Kyoto University, Kyoto
  606-8501, Japan} 

\date{\today}

\begin{abstract}
We report a systematic study of electron doping of Sr$_{2}$RuO$_{4}$ 
by non-isovalent substitution of La$^{3+}$ for Sr$^{2+}$. 
Using a combination of de Haas-van Alphen oscillations, specific heat, 
and resistivity measurements, 
we show that electron doping leads to 
a rigid-band shift of the Fermi level corresponding to one doped 
electron per La ion, 
with constant many-body quasiparticle mass 
enhancement over the band mass. 
The susceptibility spectrum is 
substantially altered and enhanced by the doping but this has 
surprisingly little effect on the strength of 
the unconventional superconducting pairing. 
\end{abstract}


\maketitle

%
The layered perovskite transition-metal-oxide Sr$_{2}$RuO$_{4}$ has 
been the subject of intensive research over the past decade. 
In its stoichiometric form, 
it can be grown with very high purity, 
allowing the observation of what is now known to be an unconventional, 
probably spin-triplet, superconducting state at low temperatures 
\cite{Maeno_Nature,Rice,Andy_Review}. 
Extensive de Haas-van Alphen (dHvA) studies have revealed precise information 
about its Fermi-surface topography with 
three nearly cylindrical sheets 
based on bands with combined Ru $d$ and oxygen $p$ character: 
one hole sheet ($\alpha$) and two electron sheets ($\beta$ and $\gamma$) 
\cite{Oguchi,Singh_Band,Andy_dHvA,Bergemann_PRL,Bergemann_Review,Damascelli_ARPES}. 
The dynamical susceptibility has features both at the wave vector 
$\textit{\textbf{q}}$ $\sim$ (2$\pi$/3,\,2$\pi$/3,\,0) \cite{Sidis}, 
and around $\textit{\textbf{q}}$\,$\sim$\,0 \cite{Braden_Sr2RuO4,Bergemann_Review}, 
both of which can be accounted for in terms of the known Fermi surfaces 
\cite{Mazin_PRL97,Bergemann_PRL,Morr}. 
In contrast to the unusually depth understanding of the normal metallic state, 
far less is known about the superconducting mechanism. 
While our experimental
and phenomenological knowledge of the superconductivity of pure
Sr$_{2}$RuO$_{4}$ is exhaustive \cite{Andy_Review}, 
it still provides insufficient constraints for models for 
the microscopic pairing mechanism. 
Here, 
the most common assumption is that the dominant sheet is $\gamma$,
since it has the largest mass-enhancement and low-$q$ susceptibility \cite{Nomura}.
One method for obtaining additional and complementary
information on correlated electron systems is
chemical doping, 
a technique that has been widely applied in recent years \cite{ARPES_Review}. 
This has motivated studies of Sr$_{2}$RuO$_{4}$ 
in which the Ru has been doped with Ti \cite{Minakata_Klaus,Braden_Ti,Kiku_PRL,Ishida_Ti} 
and Ir \cite{Kiku_JPSJ}, 
and the Sr substituted by Ca \cite{Nakatsuji}, focusing on the magnetic properties 
in each system. 
Each of these has revealed rich new physics, 
but with the commonly-experienced complication of introducing 
strong potential scattering and structural distortion. 
%

%
In this paper, 
we concentrate on non-isovalent counter-ion substitution of 
Sr$^{2+}$ with La$^{3+}$. In contrast to isovalent Ca-doping, 
the primary effect of La doping is the introduction of
extra electrons to the metallic bands at the Fermi energy. 
At the same time, the main electronic ``building blocks''---the RuO$_2$
planes---remain structurally unaffected, 
unlike the previously studied case of Ti/Ir substitution. 
Also, 
since the ionic radii of Sr$^{2+}$ and La$^{3+}$ are very similar, 
structural distortions are minimized. 
La substitution therefore provides a gentle way to study electron doping 
and the effect of changing carrier concentration in the correlated metal 
and unconventional superconductor Sr$_{2}$RuO$_{4}$.
We present here the results of a combined dHvA, resistivity, 
specific heat, and magnetic susceptibility study on samples of
Sr$_{2-y}$La$_y$RuO$_{4}$ extending up to $y$ = 0.10. 
It is remarkable in itself that the dHvA effect is observable, 
and we are able to show in detail that the normal state undergoes 
a rigid shift of the Fermi level 
with unchanged correlation quasiparticle mass enhancement: 
an unexpected result for this multi-band, correlated metal \cite{Hamacher}, 
especially considering the vicinity of the van Hove singularity 
associated with the $\gamma$ band. 
Although the absolute quasiparticle masses and the spin susceptibility spectrum 
are strongly affected by carrier doping, 
the evolution of the superconducting transition temperature $T_{\rm c}$ 
indicates no change in pairing strength,
which introduces new and strong constraints on candidate pairing mechanisms. 

%
Single crystals of Sr$_{2-y}$La$_{y}$RuO$_{4}$ with $y$ up to 0.10
were grown by a floating-zone method \cite{Mao_Crystal} with an
infrared image furnace (NEC Machinery, model SC-E15HD) at Kyoto
University.  
The La concentrations were
determined by electron-probe microanalysis (EPMA).  
Tetragonal symmetry was confirmed for all crystals by x-ray powder diffraction
measurements at room temperature.  
The lattice parameter along the in-plane direction 
increases by $\sim$\,0.2\% and that 
perpendicular to the plane decreases by $\sim$\,0.15\%
continuously up to $y$ = 0.10.
The dHvA experiments were performed at the University of St.\,Andrews
by a field modulation technique \cite{Shoenberg} at temperatures down
to 40 mK, 
with the magnetic field applied along the $c$ axis with an
accuracy of better than 3$^\circ$. 
The in-plane resistivity $\rho_{ab}$ was measured 
by a low frequency ac method between 0.3 and 5\,K. 
Magnetic susceptibility measurements were performed using 
a superconducting quantum interference device magnetometer 
(Quantum Design, MPMS-XL). 
The specific heat $C_{P}$ was measured by a thermal
relaxation method from 0.5\,K and 30\,K (Quantum Design, model PPMS).
%

\begin{figure}
\includegraphics[width=70mm]{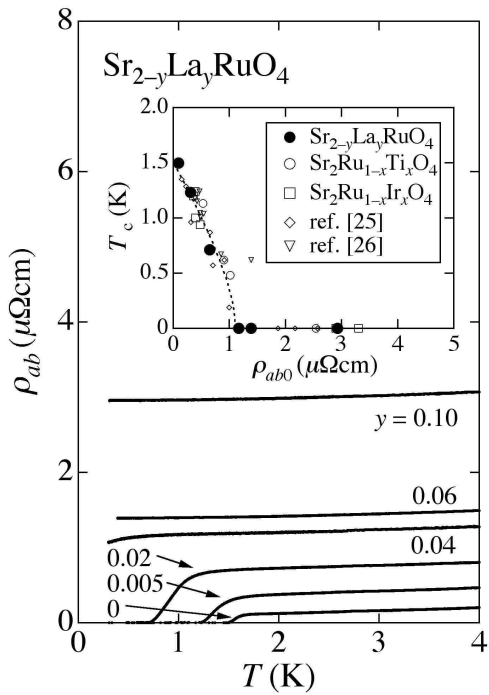}
\caption{\label{Fig1} 
Temperature dependence of the $\rho_{ab}$ in
Sr$_{2-y}$La$_{y}$RuO$_{4}$ with $y$ up to 0.10.  
Inset:
The $T_{\rm c}$ as a function of
the in-plane residual resistivity $\rho_{ab0}$ for
Sr$_{2-y}$La$_{y}$RuO$_{4}$.  
Previous results for different sources
of disorder are also shown.  
The broken line shows the Abrikosov-Gor'kov pair-breaking function. }
\end{figure}

%
Figure~\ref{Fig1} shows the temperature dependence of $\rho_{ab}$ 
in Sr$_{2-y}$La$_{y}$RuO$_{4}$ up to $y$\,=\,0.10. 
The superconducting transition temperature $T_{\rm c}$ is gradually 
and systematically suppressed with an initial rate d$T_{\rm c}$/d$y$\,$\sim$\,$-$40 K/$y$; 
it reaches zero at $y \sim 0.03$.  
As shown in the inset of Fig.~\ref{Fig1}, 
the suppression of $T_{\rm c}$ is well scaled 
by the Abrikosov-Gor'kov pair-breaking function 
for unconventional superconductivity.  
The most striking feature of the inset is that 
the data are seen to follow a universal curve 
if plotted as a function of the residual resistivity $\rho_{ab0}$ 
(defined by extrapolating the normal state $\rho_{ab}$ to $T$\,=\,0). 
$T_{\rm c}$ is completely suppressed at $\rho_{ab0}$\,$\sim$\,1.1\,$\mu\Omega$cm; 
this value and the form of the $T_{\rm c}(\rho_{ab0})$ curve are identical 
for La and various other kinds of impurities and defects 
\cite{Andy_Defect,Mao_Defect,Kiku_PRL,Kiku_JPSJ}.
The rate at which the La dopants between the RuO$_{2}$ planes introduce
scattering is, 
as expected for an out-of-plane dopant, 
much smaller than that for in-plane substituted
impurities such as Ti and Ir for Ru \cite{Kiku_PRL,Kiku_JPSJ}. 
The residual resistivity, 
$\rho_{ab0}$, increases systematically with $y$ 
at the rate of d$\rho_{ab0}$/d$y$\,$\sim$\,40\,$\mu\Omega$cm/$y$, 
that is, with a phase shift for impurity scattering $\delta_{0}$\,$\sim$\,$\pi/12$. 
In contrast, 
Ti and Ir act as unitary scatterers with $\delta_{0}$\,$\sim$\,$\pi$/2 \cite{Kiku_JPSJ}. 
However, we reiterate that 
although the rate at which the La ions affect the resistivity is lower than for Ti or Ir, 
the effect of that change of resistivity on the
superconductivity is independent of the dopant species. 
%
 

\begin{figure}
\includegraphics[width=70mm]{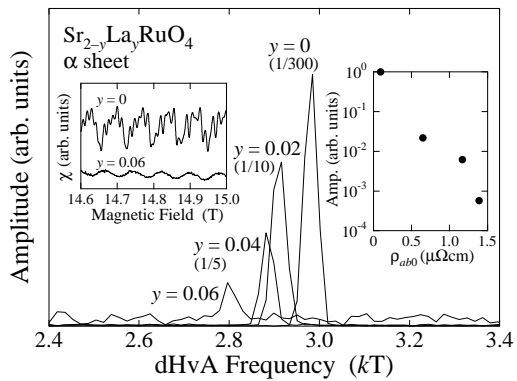}
\caption{\label{Fig2} 
The dHvA frequency spectrum for the $\alpha$ sheet of
Sr$_{2-y}$La$_{y}$RuO$_{4}$ up to $y$\,=\,0.06. 
The left inset shows quantum oscillatory susceptibility $\chi$ of
Sr$_{2-y}$La$_{y}$RuO$_{4}$ with $y$\,=\,0 and 0.06. 
The dependence of the dHvA amplitude on the in-plane residual resistivity
$\rho_{ab0}$ is presented in the right inset. 
Note the logarithmic scale on the vertical axis.}
\end{figure}

\begin{figure*}
\includegraphics[width=150mm]{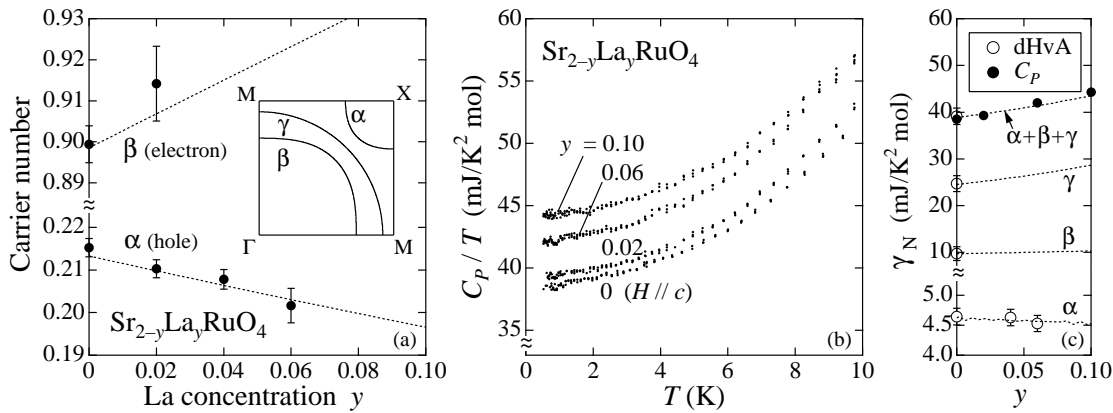}
\caption{\label{Fig3} 
(a) La concentration dependence of the carrier number of 
the $\alpha$ and the $\beta$ sheets. 
The $\gamma$ sheet dependence is
not shown since we have data only at $y$\,=\,0, 
but it corresponds to an increase of approximately 0.03 electrons by $y$\,=\,0.06.  
Inset: A sketch of the Fermi surface of Sr$_{2}$RuO$_{4}$. 
(b) Temperature dependence of $C_{P}/T$ in 
Sr$_{2-y}$La$_{y}$RuO$_{4}$ up to $y$\,=\,0.10. 
(c) Doping dependence of $\gamma_{\rm N}$ (for definition see text). 
Closed (open) circles represent data from specific heat (dHvA) measurements. 
The dotted lines show the prediction of the calculation described in the text. 
The error bars on the dHvA-derived sheet-specific contributions reflect 
the combined uncertainties of the data and the parameters used in the calculation.}
\end{figure*}

%
Fig.~\ref{Fig2} shows the Fourier transform of the dHvA oscillations
for Sr$_{2-y}$La$_{y}$RuO$_{4}$ up to $y$\,=\,0.06, 
in a narrow frequency region around the $\alpha$ branch. 
Sample oscillations are
shown in the left inset of Fig.~\ref{Fig2} for $y$\,=\,0 and 0.06.  
For pure Sr$_{2}$RuO$_{4}$ with $T_{\rm c}$\,=\,1.44 K, 
not only the whole frequency spectrum with three Fermi-surface sheets 
but also additional harmonics $\alpha$ 
and linear combinations such as $\alpha$\,+\,$\beta$ are detected.  
The oscillatory amplitude is exponentially 
suppressed by La substitution, 
reflecting the introduction of weak disorder, 
as seen in the right inset in Fig.~\ref{Fig2}.  
The effect is strongest on the large $\gamma$ sheet so that even at $y$\,=\,0.02
($\rho_{ab0}$\,$\sim$\,0.7\,$\mu\Omega$cm) its signal is unobservable. 
For $y$\,=\,0.06, only the $\alpha$ frequency remains detectable 
(left inset of Fig.~\ref{Fig2}), 
and no oscillations at all are found for $y$\,=\,0.10 in our current study. 
The exponential decay of the signal with doping gives the opportunity
for a reliable Dingle analysis.  
The suppression factor is $\exp(-\pi r_c/\ell)$, 
where $r_{c}$ is the cyclotron radius and 
$\ell$ is the carrier mean free path. 
The values obtained for the $\alpha$ sheet
range from 990$\pm$100 nm for $y$\,=\,0 to 76$\pm$9 nm for $y$\,=\,0.06. 
These agree, within the stated error, with the values obtained by an
analysis of the resistivity under the ``isotropic mean-free-path approximation'' 
\cite{Andy_dHvA,Bergemann_Review}.  
This is interesting 
because it implies that the resistivity, 
which is biased towards large angle scattering, 
still gives a good estimate of the scattering (including small angle events) that 
damps the quantum oscillations, 
even though the out-of-plane La ions act as relatively diffuse scattering centers. 
%

%
A more significant feature of the data is that 
we observe progressive changes in the dHvA oscillation frequencies $F_{\rm ext}$, 
as shown in Fig.~\ref{Fig2}. 
Since these frequencies directly relate to the cross-sectional areas of 
the Fermi-surface sheets via
$A_{\rm ext} = 2\pi eF_{\rm ext}/\hbar$, 
this reflects a change in the carrier concentration associated with each sheet.  
As seen in Fig.~\ref{Fig3}(a), 
the hole-like $\alpha$ sheet shrinks 
while the electron-like $\beta$ sheet grows in size, 
which is completely consistent with the La acting as an electron donor.
A much more quantitative analysis is also possible in this system. 
The years of work on pure Sr$_{2}$RuO$_{4}$ 
have led to the construction of an empirically determined tight-binding model 
for the electronic structure, 
derived from fits to the experimentally determined Fermi surface \cite{Bergemann_Review}. 
The dotted lines in Fig.~\ref{Fig3} are not fits or guides to the eye, 
but topography calculations based on this model, 
without free parameters, 
under the assumption that each La dopes one free electron
and induces a rigid shift of the Fermi level.  
As can be seen, 
the agreement is excellent. 
This is remarkable, 
because although Luttinger's theorem usually puts strong enough constraints on
the Fermi-surface geometry for such rigid-band shift calculations to
work at least approximately in single-band systems, 
the constraints are much less powerful for a multi-band system such as Sr$_2$RuO$_{4}$. 
In Ca-substituted Sr$_2$RuO$_{4}$,
for example, 
doping lowers the $d_{xy}$ band with respect to the $d_{yz/zx}$ system \cite{Nakatsuji}; similarly, 
correlations among electrons can influence the evolution of the Fermi-surface geometry, 
especially in metals with both electron and hole pockets \cite{Hamacher} 
---but in strongly correlated Sr$_{2-y}$La$_{y}$RuO$_{4}$, 
both effects are either absent or compensate each other. 
%


%
We can also use the tight-binding model and its density of states  
to predict each sheet's contribution to the specific heat $C_{P}/T$ 
as a function of $y$. 
The experimental $C_{P}/T$ is strongly enhanced
over the bare density of states by electron-phonon and
electron-electron interactions; 
we empirically set this enhancement to be independent of $y$ 
and use the well-known values for pure Sr$_2$RuO$_{4}$ \cite{Bergemann_Review}.
In Fig.~\ref{Fig3}(b), 
we show the $C_{P}/T$ as a function of temperature for various values of $y$. 
The phonon term has not been extracted, 
so the electronic contribution, 
defined as $\gamma_{\rm N}$, 
is given by the extrapolation to zero temperature 
(well approximated by the lowest temperature value in each case). 
These values are plotted against $y$ in Fig.~\ref{Fig3}(c) (filled circles); 
that plot also contains the individual sheets' contributions (open circles) 
as inferred from analysis of dHvA temperature damping. 
The dashed lines
are the calculations from the tight-binding model, 
in which the quasiparticle mass-enhancement
over the band mass is taken to be constant throughout the doping range,  
and again, 
the agreement is excellent: 
the experimental $C_{P}/T$ rises with $y$, 
and the contribution from the $\alpha$ sheet decreases slightly, 
in line with the calculation. 
%

%
The rapidly increasing experimental value for the electronic
$C_{P}/T$, and the underlying tight-binding model, 
imply a large change ($>$\,15\%) to the $\gamma$ sheet mass on doping.  
This is not surprising, 
since the electron doping is qualitatively expected to
shift the Fermi level for that band towards the van Hove singularity 
\cite{Singh_Band,Bergemann_Review,Nomura}.  
We have observed an even 
more substantial (30\%) change to the low temperature static susceptibility $\chi$($q$\,=\,0)  
(data not shown): 
the increase in the density of states is augmented here 
by the feedback mechanism arising from the Stoner factor. 
%

%
Along with the bulk $\chi$($q$\,=\,0), 
the whole spin-susceptibility spectrum $\chi(\textit{\textbf{q}})$ 
has to change significantly on electron doping. 
In pure Sr$_{2}$RuO$_{4}$, 
$\chi(\textit{\textbf{q}})$ reflects the nesting
properties of the Fermi surface \cite{Mazin_PRL97,Bergemann_PRL,Morr},
with features both at wave vector $\textit{\textbf{q}}$ $\sim$
(2$\pi$/3,\,2$\pi$/3,\,0) \cite{Sidis} from $\alpha/\beta$ nesting, and
at low $q$ \cite{Braden_Sr2RuO4,Bergemann_Review} from the
$\gamma$ sheet \cite{Ng}. 
Electron doping will shift the $\alpha/\beta$ nesting to smaller wave vectors, 
and enhance the low-$q$ susceptibility as the $\gamma$ sheet 
moves closer to the van Hove singularity. 
%

%
Many theories attribute the superconducting pairing in Sr$_{2}$RuO$_{4}$ 
to spin fluctuations in either the $\gamma$ or $\alpha/\beta$ channel, 
where the more common assumption is that the dominant sheet is $\gamma$, 
since it has the largest mass-enhancement and low-$q$ susceptibility. 
The relation between the spin-fluctuation spectrum $\chi(\textit{\textbf{q}})$
and $T_{\rm c}$ and $\xi$ is a subtle one and certainly 
beyond the scope of this paper. 
There are indications that in Sr$_{2}$RuO$_{4}$ 
the low value of $T_{\rm c}$, 
especially when compared with the cuprate high-$T_{\rm c}$ superconductors, 
is due to competition and
near-cancellation between two different pairing symmetries
\cite{Monthoux_Private}. 
It is to be expected, 
then, 
that changes in $\chi(\textit{\textbf{q}})$ would have drastic effects on 
$T_{\rm c}$ and coherence length $\xi$ 
which should be visible as deviations from the universal
Abrikosov-Gor'kov curve in the inset of Fig.~\ref{Fig1}.
Within experimental errors, 
and for the doping range in which 
we were able to establish the superconducting properties, 
we see no such deviations. 
We therefore believe that our observations place
significant constraints on the search for the mechanism of the
superconductivity of Sr$_{2}$RuO$_{4}$. 
In this context, 
it would also be interesting to directly measure 
the dynamical susceptibility of these samples, 
to gauge the extent to which the spin-fluctuation
spectrum is changing as a function of $\textit{\textbf{q}}$.
%


%
In summary, 
we have studied the microscopic effects of doping
La$^{3+}$ for Sr$^{2+}$ in the correlated electron metal Sr$_{2}$RuO$_{4}$. 
An empirical, 
rigid-band shift, 
tight-binding
parameterization of the electronic structure that 
incorporates constant many-body renormalizations 
allows a quantitative prediction of the evolution of 
both the Fermi surface geometry and the thermal properties of the doped material.  
Although this is not the first time
that dHvA has been observed in the presence of doping in a correlated
electron metal 
(a notable previous example is Ce$_{1-x}$La$_{x}$B$_{6}$ \cite{Goodrich}), 
it is to our knowledge 
the first time that the rigid-band model has been put to such a sensitive test, 
especially for multi-band system. 
The superconducting properties remain remarkably unchanged 
in the face of a rapidly evolving, 
enhanced susceptibility spectrum,
which raises intriguing questions about the mechanism of the
unconventional superconductivity.
%


The authors thank A.J. Millis, T. Nomura, M. Braden, Kosaku
Yamada, P. Monthoux, and G.G. Lonzarich for useful
discussions.  They also thank H. Fukazawa for technical supports and
discussions, M. Yoshioka for technical supports, Y. Shibata and
Takashi Suzuki for EPMA measurements at Hiroshima University.
This work was in part supported by the Grant-in-Aid for Scientific Research 
(S) from the Japan Society for Promotion of Science (JSPS), by the Grant-in-Aid 
for Scientific Research on Priority Area 'Novel Quantum Phenomena in 
Transition Metal Oxides' from the Ministry of Education, Culture, Sports, 
Science and Technology, and by the Leverhulme Trust.
One of the authors (N.K.) is supported by JSPS Postdoctoral Fellowships 
for Research Abroad, 
while S.A.G. gratefully acknowledges the support of the Royal Society.

%


\begin{thebibliography}{99}

\bibitem{Maeno_Nature}
Y. Maeno $et~al$., Nature \textbf{372}, 532 (1994). 

\bibitem{Rice}
T.M. Rice and M. Sigrist, J. Phys. Condens. Matter \textbf{7}, L643 (1995). 

\bibitem{Andy_Review}
A.P. Mackenzie and Y. Maeno, Rev. Mod. Phys. \textbf{75}, 657 (2003). 

\bibitem{Andy_dHvA}
A.P. Mackenzie $et~al$., Phys. Rev. Lett. \textbf{76}, 3786 (1996). 

\bibitem{Bergemann_PRL}
C. Bergemann $et~al$., Phys. Rev. Lett. \textbf{84}, 2662 (2000). 

\bibitem{Bergemann_Review}
C. Bergemann $et~al$., Adv. Phys. \textbf{52}, 639 (2003). 

\bibitem{Oguchi}
T. Oguchi, Phys. Rev. B. \textbf{51}, 1385 (1995). 

\bibitem{Singh_Band}
D.J. Singh, Phys. Rev. B. \textbf{52}, 1358 (1995). 

\bibitem{Damascelli_ARPES}
A. Damascelli $et~al$., Phys. Rev. Lett. \textbf{85}, 5194 (2000). 

\bibitem{Sidis}
Y. Sidis $et~al$., Phys. Rev. Lett. \textbf{83}, 3320 (1999). 

\bibitem{Braden_Sr2RuO4}
M. Braden $et~al$., Phys. Rev. B. \textbf{66}, 064522 (2002). 

\bibitem{Mazin_PRL97}
I.I. Mazin and D.J. Singh, Phys. Rev. Lett. \textbf{79}, 733 (1997). 

\bibitem{Morr}
D.K. Morr P.F. Trautman, and M.J. Graf, Phys. Rev. Lett. \textbf{86}, 5978 (2001). 

\bibitem{Nomura}
T. Nomura and K. Yamada, J. Phys. Soc. Jpn. \textbf{69}, 3678 (2000); 
$ibid$. \textbf{71}, 1993 (2002). 

\bibitem{ARPES_Review}
For instance, 
A. Damascelli, Z. Hussain, and Z.X. Shen, Rev. Mod. Phys. \textbf{75}, 473 (2003); 
J.M.D. Coey, M. Viret, and S.v. Molnar, Adv. Phys. \textbf{48}, 167 (1999). 

\bibitem{Minakata_Klaus}
M. Minakata and Y. Maeno, Phys. Rev. B \textbf{63}, 180504(R) (2001); 
K. Pucher $et~al$., Phys. Rev. B \textbf{65}, 104523 (2002). 

\bibitem{Braden_Ti}
M. Braden $et~al$., Phys. Rev. Lett. \textbf{88}, 197002 (2002).

\bibitem{Kiku_PRL}
N. Kikugawa and Y. Maeno, Phys. Rev. Lett. \textbf{89}, 117001 (2002). 

\bibitem{Ishida_Ti}
K. Ishida $et~al$., Phys. Rev. B \textbf{67}, 214412 (2003). 

\bibitem{Kiku_JPSJ}
N. Kikugawa, A.P. Mackenzie, and Y. Maeno, J. Phys. Soc. Jpn. \textbf{72}, 237 (2003). 

\bibitem{Nakatsuji}
S. Nakatsuji and Y. Maeno, Phys. Rev. Lett. \textbf{84}, 2666 (2000). 

\bibitem{Hamacher}
K. Hamacher, C. Gros, and W. Wenzel, Phys. Rev. Lett. \textbf{88},
217203 (2002).

\bibitem{Mao_Crystal}
Z.Q. Mao, Y. Maeno, and H. Fukazawa, Mat. Res. Bull. \textbf{35}, 1813 (2000). 

\bibitem{Shoenberg}
D. Shoenberg, \textit{Magnetic Oscillations in Metals} 
(Cambridge University Press, Cambridge, 1976). 

\bibitem{Andy_Defect}
A.P. Mackenzie $et~al$., Phys. Rev. Lett. \textbf{80}, 161 (1998). 

\bibitem{Mao_Defect} 
Z.Q. Mao, Y. Mori, and Y. Maeno, Phys. Rev. B. \textbf{60}, 610 (1999).

\bibitem{Ng}
K. Ng and M. Sigrist, J. Phys. Soc. Jpn. \textbf{69}, 3764 (2000). 

\bibitem{Monthoux_Private} 
P. Monthoux and G.G. Lonzarich, private communication.

\bibitem{Goodrich}
R.G. Goodrich $et~al$., Phys. Rev. Lett. \textbf{82}, 3669 (1999). 




\end{thebibliography}
\end{document}